\documentclass[11pt,superscriptaddress, onecolumn, nofootinbib]{revtex4-2}
\usepackage{graphicx}
\usepackage{amssymb,amsmath}
\usepackage{natbib}
\begin{document}
\title{The Gauge issue and the Hamiltonian theory
	of
	cosmological perturbations
\thanks{Presented at POTOR8}%
}
\author{Alice Boldrin}
\email{Alice.Boldrin@ncbj.gov.pl}
\affiliation{National Centre for Nuclear Research, Pasteura 7, 02-093
	Warszawa, Poland}

\begin{abstract}
	We present a general formalism for the Hamiltonian description of perturbation theory around any spatially homogeneous spacetime. We employ and refine the Dirac method for constrained systems, which is very well-suited to cosmological perturbations. This approach includes a discussion of the gauge-invariant dynamics of perturbations as well as an analysis of gauge transformations, gauge-fixing, partial gauge-fixing and spacetime reconstruction. We will introduce the Kucha\v r parametrization of the kinematical phase space as a convenient tool for studying the gauge transformations. The key element of this approach is the reconstruction of spacetime based on gauge-fixing conditions.
\end{abstract}

\maketitle

\section{Introduction}
In the attempt to obtain a quantum theory suitable for the description of the primordial structure of the Universe, we study the Hamiltonian formalism for cosmological perturbation theory (CPT). This work has been done before with different background spacetime models like the Friedman universe \cite{Malkiewicz:2018ohk} and the Bianchi Type I model \cite{Boldrin:2021xrm}. Our aim is to study the complete Hamiltonian formalism in a general background focusing on the gauge independent description of CPT as well as the issue of gauge fixing (see e.g. \cite{Malik:2012dr, Kodama:1984ziu} for alternative discussions on the gauge issue in CPT), gauge transformations and spacetime reconstruction.
We employ the Dirac method \cite{Dirac} to study the Hamiltonian in different gauges and reconstruct the spacetime metric from gauge-invariant quantities (Dirac observables). We also discuss an alternative method based on the so-called Kucha\v r decomposition \cite{Kuchar} which provides a parametrization of the phase space in which the constrains play the role of canonical variables conjugate to the gauge-fixing conditions. For a more detailed discussion and for an application of the presented method, see \cite{Boldrin:2022vcp}.

\section{Cosmological perturbation theory}

The Hamiltonian in the the Arnowitt-Deser-Misner (ADM) formalism \cite{ADM} expanded to second order reads\footnote{We assume the topology of the spacetime to be $\mathcal{M}\simeq \mathbb{T}^3\times \mathbb{R}$ so to have a spatially compact universe and avoid ambiguous definitions of the symplectic structure for background (homogeneous) variables. }
\begin{equation}\label{totham}
	\mathbb{H}=
	\int_{\mathbb{T}^3}
	\left(
	\overline{N}\mathcal{H}_0^{(0)}+
	\overline{N}\mathcal{H}^{(2)}_0+
	\delta N^\mu\delta\mathcal{H}_\mu
	\right)d^3x
\end{equation}
where $\overline{N}$ is the background lapse function and $\delta N^\mu$, with $\mu=0,i$, are the first order lapse and shift functions. The Hamiltonian densities $\mathcal{H}^{(0)}$ and $\mathcal{H}^{(2)}$ are respectively zeroth  and second order, whereas  $\delta\mathcal{H}_\mu$ represent the first order constraints. 
We assume a spatially homogeneous background spacetime with spatial coordinates defined such that the background shift vector $N^i$ vanishes as well as the background Hamiltonian $\mathcal{H}_i^{(0)}$.
The Hamiltonian \eqref{totham} is a function of the background canonical variables $\bar{q}_{ij}$ and $\bar{\pi}_{ij}$ which are respectively the three-metric and three-momenta, and the perturbed variables defined as 
$	\delta q_{ij}=q_{ij}-\overline{q}_{ij}$ and	$\delta\pi^{ij}=\pi^{ij}-\overline{\pi}^{ij}$.

The Hamiltonian \eqref{totham} defines a gauge for the following reasons:

First, at each spatial point the constraints algebra is closed, i.e.
\begin{align}
	\{\delta\mathcal{H}_i,\delta\mathcal{H}_j\}=0,
	\quad
	\{\delta\mathcal{H}_0,\delta\mathcal{H}_i\}=0,	
\end{align}
where this result is true for any homogeneous background.
Furthermore the constraints are dynamically stables, i.e.
\begin{align}\label{hypalg}
	\{\mathbb{H},\delta\mathcal{H}_0\}=-\delta\mathcal{H}^i_{,i}(x)\approx0,
	\quad
	\{\mathbb{H},\delta\mathcal{H}_i\}=0,	
\end{align}
where the "weak equality" $\approx$, means that the equality holds in the constraint surface.
\section{Gauge-Fixing and Dirac procedure}
\begin{figure}[htb]
	\centerline{%
		\includegraphics[width=12.5cm]{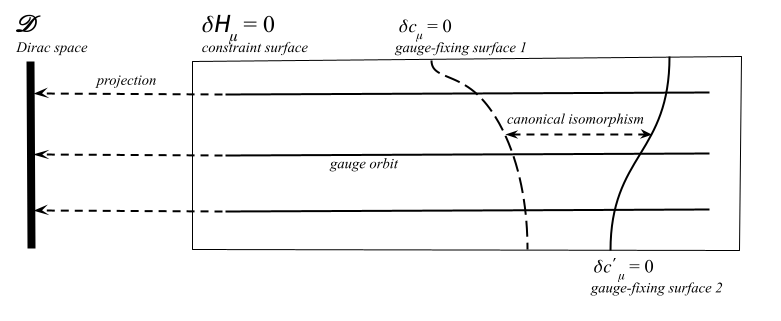}}
	\caption{Graphical representation of the Dirac procedure.}
	\label{Fig:F2H}
\end{figure}

The four constraints $\delta\mathcal{H}_\mu$ generate a gauge freedom which can be removed by imposing four gauge-fixing conditions $\delta c_\mu=0$. The Poisson bracket between the gauge-fixing conditions and the constraints form an invertible matrix $\det\{\delta c_\mu,\delta\mathcal{H}_\mu\}\neq 0$.
Applying the constraints and the gauge-fixing conditions we can reduce our Hamiltonian which will now depend on 4 physical variables $(\delta q_I^{phys},\delta \pi^I_{phys})$ instead of the 12 ADM perturbation\footnote{We assume the vacuum case for the sake of clarity. See \cite{Malkiewicz:2018ohk} or \cite{Boldrin:2021xrm} for the Dirac method applied when there is matter content.} variables $(\delta q_{ij},\delta \pi^{ij})$. Those new variables form a canonical coordinate system on the submanifold in the kinematical phase space. This submanifold is thus called the physical phase space\footnote{It's canonical structure is now given by the Dirac brackets $\{.,.\}_D =\{.,.\}-\{.,\delta \phi_\mu\}\{\delta \phi_\mu,\delta \phi_\nu\}^{-1}\{\delta \phi_\nu,.\}$, where $\delta \phi_\mu\in(\delta\mathcal{H}_\mu,\delta c_\mu)$. }.
The parametrization provided by these physical variables is defined by the gauge-fixing surface that intersects all gauge orbits (see Fig. \ref{Fig:F2H} ).
It is convenient to define a set of gauge-independent variables defined as \begin{align}\label{diracobs}
	\{\delta D_I,\delta \mathcal{H}_\mu\}\approx 0, \, \forall \mu,
\end{align}
which parametrize the space of gauge orbits in the constraints surface. Those variables are known as Dirac observables and are equal to the number of physical variables. There exists a one-to-one correspondence between the Dirac observables and the physical variables, such that 
\begin{align}
	\delta D_I+\epsilon_I^\mu\delta c_\mu+\xi^\mu_I \delta \mathcal{H}_\mu=\delta O^{phys}_I(\delta q^{phys}_I,\delta \pi^I_{phys})
\end{align}
where $\epsilon_I^\mu$ and $\xi^\mu_I$ are background coefficients.
Using this new parametrization the Hamiltonian can be written in a gauge-independent manner as $\mathcal{H}_{phys}^{(2)}=\mathcal{H}_{red}^{(2)}+\mathcal{H}_{ext}^{(2)}$, where $\mathcal{H}_{phys}^{(2)}$ denotes the so called physical Hamiltonian, $\mathcal{H}_{red}^{(2)}$ is the reduced Hamiltonian in terms of the physical variables and $\mathcal{H}_{ext}^{(2)}$ is the extra Hamiltonian generated by the time-dependent canonical transformation needed to change parametrization. 
\section{Spacetime reconstruction}\label{spaserec}
In the previous section we discussed how to obtain the physical Hamiltonian. In order to reconstruct the spacetime we still need to find the values of the first-order lapse and shift.
To do so we use the consistency equation $\{\delta c_\mu,\mathbb{H}\}=0$, which, from Eq. \eqref{totham}, implies
\begin{align}
	\frac{\delta N^\mu}{N}=-\{\delta c_\nu,\delta \mathcal{H}_\mu\}^{-1}\left(\{\delta c_\nu,\delta \mathcal{H}^{(0)}\}+\{\delta c_\nu,\mathcal{H}^{(2)}\}\right)
\end{align}
This equation is only meaningful in the constraint surface.
\section{Kucha\v r decomposition}
We present a different parametrization of the kinematical phase space where the constraints take the role of canonical variables.
For instance, we define two sets of canonical variables. The first set comprises the  first order constraints $\delta \mathcal{H}_\mu$ and the 4 gauge-fixing functions, here denoted as $\delta C^\mu$. The second pair of canonical variables is given by the Dirac observables $\delta D_I$, defined in Eq.\eqref{diracobs}.
The Hamiltonian written in this parametrization will then be
\begin{align}\label{kucham}
	\mathbb{H}\rightarrow \mathbb{H}_K=\mathbb{H}+\mathbb{K}
	=\int 
		\left(
	\overline{N}\mathcal{H}_0^{(0)}+
	\overline{N}(\mathcal{H}^{(2)}_0+\mathcal{K})+
	\delta N^\mu\delta\mathcal{H}_\mu
	\right)d^3x
	,
\end{align}
where $\mathbb{K}$ is the extra Hamiltonian coming from the time-dependent parametrization.
We notice that, since the constraints are conserved in the constraint surface, terms of the form $\propto \delta C^\mu \delta C^\nu$, $\propto \delta Q_I \delta C^\mu$ and $\propto \delta P_I \delta C^\mu$ are not present in Eq. \eqref{kucham}.
Moreover, considering that $\mathcal{H}^{(2)}\approx \mathcal{H}^{(2)}_{red}$ and $\mathcal{K}^{(2)}\approx \mathcal{H}^{(2)}_{ext}$, which tells us the two dynamics must be weakly equal, we have that the total Hamiltonian can only be of the form
\begin{align}\label{TOTkucham}
	\mathbb{H}_K
	=
	&N
	\int 
	\bigg[
\underbrace{	\mathcal{H}^{(2)}_{phys}(\delta Q_I,\delta P^I)}_{\text{physical part}}+
	\nonumber\\
&
+
\underbrace{	\bigg(
	\lambda^{\mu I}_1\delta Q_I
	+
	\lambda^{\mu }_{2I}\delta P_I
		+
	\lambda^{\mu \nu}_{3}\delta \mathcal{H}_\nu
		+
	\lambda^{\mu }_{4\nu}\delta C^\nu
	+\frac{\delta N^\mu}{N}
	\bigg)
	\mathcal{H}_\mu}_{\text{weakly vanishing part}}
	\bigg]d^3x
	,
\end{align}
where $\lambda^{\mu I}_1$, $\lambda^{\mu }_{I2}$ and $\lambda^{\mu \nu}_3$ are zeroth-order coefficients that can depend on the gauge-fixing $\delta C^\mu$. The value of $\lambda^{\mu}_{4\nu}$ is gauge-invariant, it is showed to be fixed unambiguously by the algebra of the hypersurface \eqref{hypalg}.
 
 \subsection{Gauge transformations}
 An interesting property of the Kucha\v r decomposition comes from the freedom in the choice of the canonical variable $\delta C^\mu$. This means that we have a class of parametrizations of the kinematical phase space. In particular, we can define the new set of gauge-fixing conditions as $\delta \tilde{C}^\mu$, and the full gauge transformation will be given by the map $\mathbb{G}:(\delta \mathcal{H}_\mu,\delta C^\mu,\delta Q_I,\delta P^I)\rightarrow (\delta \tilde{\mathcal{H}}_\mu,\delta \tilde{C}^\mu,\delta \tilde{Q}_I,\delta \tilde{P}^I)$, where $\delta\mathcal{H}_\mu=\delta\tilde{\mathcal{H}}_\mu$.
 We are free to assume that the new gauge-fixing functions are thus canonically conjugate to the constraints $\delta\mathcal{H}_\mu$. Thus we have $\{\delta\mathcal{H}_\nu,\delta \tilde{C}^\mu-\delta C^\mu\}=0$, which implies 
 \begin{align}\label{ctilde}
 	\delta \tilde{C}^\mu
 	=\delta {C}^\mu+\alpha^\mu_I \delta P^I+\beta^{\mu I} \delta Q_I+\gamma^{\mu\nu}\delta\mathcal{H}_\nu,
 	\end{align}
 where $\alpha^\mu_I$, $\beta^{\mu I}$ and $\gamma^{\mu\nu}$ are background parameters.
 
The gauge-fixing condition is only relevant in the constraints surface, so Eq.\eqref{ctilde} is fully determined by the parameters $\alpha^\mu_I$ and $\beta^{\mu I}$. Moreover it means that the space of gauge-fixing conditions is the affine space of dimension equal to the number od Dirac observables. The introduction of a different gauge will lead to the new Hamiltonian $\mathbb{H}_{\tilde{K}}$, with an extra Hamiltonian density $\Delta \mathcal{K}^{(2)}$. Studying the new symplectic form of the system we find that $\gamma^{\mu\nu}$ depends only on $\alpha^\mu_I$ and $\beta^{\mu I}$, which thus are the only parameters needed to uniquely determine the gauge transformation.
\subsection{Spacetime reconstruction}
As discussed in Sec. \ref{spaserec}, the spacetime reconstruction is obtained by the dynamical equations $\delta \dot{C}^\mu=0$, which means that it is sensitive to the chosen parametrization.
In particular in the Kucha\v r parametrization we will have $\{\delta C^\nu,\mathbb{H}_{K}\}_K=0$, which, from Eq. \eqref{kucham} becomes
\begin{align}
	\frac{\delta N^\mu}{N}=-\frac{\partial (\mathcal{H}^{(2)}+\mathcal{K}^{(2)})}{\partial \delta \mathcal{H}_\mu}.
\end{align}
Notice the above formula only depends on the weakly vanishing part of the Hamiltonian since the lapse and shift are gauge-dependent quantities.
It is interesting to consider the difference between the lapse and shift in two gauges. Using Eq. \eqref{TOTkucham}, we have
\begin{align}\label{lapseshift}\begin{split}
		&\frac{\delta\tilde{N}^\mu}{N}\bigg|_{\delta\tilde{C}^{\mu}=0}- \frac{\delta{N}^\mu}{N}\bigg|_{\delta{C}^{\mu}=0}
		\approx
		\\
		&\quad
		\approx 
		\left(
		\lambda_{4\nu}^\mu\beta^{\nu I}+\dot{\beta}^{\mu I}+\frac{\partial^2\mathcal{H}^{(2)}_{phys} }{\partial \delta Q_I\partial \delta P^J}\beta^{\mu J}-\frac{\partial^2\mathcal{H}^{(2)}_{phys} }{\partial \delta Q_I\partial \delta Q_J}\alpha^\mu_{~J}
		\right)\delta{Q}_I\\
		&\quad
		+
		\left(
		\lambda_{4\nu}^\mu\alpha^\nu_{~I}+\dot{\alpha}^\mu_{~I}-\frac{\partial^2\mathcal{H}^{(2)}_{phys} }{\partial \delta P^I\partial \delta Q_J}\alpha^\mu_{~J}+\frac{\partial^2\mathcal{H}^{(2)}_{phys} }{\partial \delta P^I\partial \delta P^J}\beta^{\mu J}
		\right)\delta{P}^I.
\end{split}\end{align}
We see that the spacetime reconstruction in a new gauge can be obtained by the lapse and shift in the initial gauge plus some terms which solely depend on the physical part of the Hamiltonian $\mathcal{H}^{(2)}_{phys}$ and the gauge-invariant coefficient $\lambda^\mu_{4\nu}$, which can be obtained from the algebra of the hypersurface deformations.
\section{Partial gauge-fixing}
We previously discussed the gauge-fixing defined as setting the conditions $\delta C^\mu=0$. However it can be interesting to study the case in which these 4 conditions are substituted with conditions on the lapse and shift functions. This is what we call partial gauge-fixing.
From this consideration we can study the transformations which preserve the lapse and shift functions, that is, $\frac{\delta\tilde{N}^\mu}{N}\big|_{\delta\tilde{C}^{\mu}=0}- \frac{\delta{N}^\mu}{N}\big|_{\delta{C}^{\mu}=0}=0$.
Using Eq. \eqref{lapseshift} and solving it for $\alpha^\nu_{~I}$ and $\beta^{\mu I}$, we can solve the ambiguity in the choice of the gauge-fixing condition.
\begin{align}\label{kernel}\begin{split}
		\dot{\alpha}^\mu_{~I} &=-
		\beta^{\mu J}
		\frac{\partial^2\mathcal{H}^{(2)}_{phys} }{\partial \delta P^J\partial \delta P^I}
		+\alpha^\mu_{~J}
		\frac{\partial^2\mathcal{H}^{(2)}_{phys} }{\partial \delta Q_J\partial \delta P^I}-\lambda_{4\nu}^\mu\alpha^\nu_{~I},\\
		\dot{\beta}^{\mu I} &=
		-\beta^{\mu J}
		\frac{\partial^2\mathcal{H}^{(2)}_{phys} }{\partial \delta P^J\partial \delta Q_I}
		+\alpha^\mu_{~J}
		\frac{\partial^2\mathcal{H}^{(2)}_{phys} }{\partial \delta Q_J\partial \delta Q_I}-\lambda_{4\nu}^\mu\beta^{\nu I},
	\end{split}
\end{align}
The above equations fix the gauge-fixing function at all times once $\delta C^\mu(t_0)$ is fixed at an initial time $t_0$. This means that the choice of $\delta C^\mu(t_0)$ fixes the initial three-surface. Given the initial values of the Dirac observables $(\delta Q_I(t_0),\delta P^I(t_0))$, we are able to explicitly reconstruct the initial three-surface  in terms of the ADM perturbation variables.
Moreover we are able to fully reconstruct the spacetime geometry since the evolution of the three-surface with its coordinates is completely determined by the evolution of the gauge-fixing function $\delta \tilde{C}^\mu(t)$ and the independent evolution of the gauge-invariant variables\footnote{The spacetime coordinates system is independent from the evolution of this variables.} $(\delta Q_I(t),\delta P(t))$.

\section{Conclusions}
We were able to simplify the Hamiltonian approach to CPT by showing that it is possible to separate the gauge-independent dynamics of perturbation from the issues of gauge-fixing and spacetime reconstruction. In particular we showed how the spacetime reconstruction can be pursued with the sole knowledge of gauge-fixing conditions. Moreover the discussed Kuca\v r decomposition serves as a useful and insightful tool to the study of gauge-fixing conditions and spacetime reconstruction. The space of gauge-fixing conditions and the formula for the spacetime reconstruction is given explicitly for any gauge.

This approach might be applied to multiple conceptual problems in quantum cosmology, such as the time problem, the semi-classical spacetime reconstruction , or the relation between the kinematical and reduced phase space quantization.
Moreover, the complete control over the gauge-fixing issue provided by the presented method, could be very useful for the problem of gluing perturbed spacetimes to other spacetime models (e.g., ones that include non-linearities). The choice of the gluing surface and its coordinates should be nicely described by our method.
\section*{Acknowledgments}
	The author acknowledge the support of the National Science Centre (NCN, Poland) under the research grant 2018/30/E/ST2/00370.



\end{document}